\documentclass{WileyMSP-template}

\usepackage{graphicx}
\usepackage{amsmath}

\usepackage[usenames,dvipsnames]{xcolor}

\newcommand{\rf}[1]{(\ref{#1})}

\newcommand{\beq}{\begin{equation}}
\newcommand{\eeq}{\end{equation}}
\newcommand{\beqr}{\begin{eqnarray}}
\newcommand{\eeqr}{\end{eqnarray}}
\newcommand{\lb}[1]{\label{#1}}
\newcommand{\bc}{\begin{center}}
\newcommand{\ec}{\end{center}}
\newcommand{\ct}[1]{\cite{#1}}

\begin{document}

\pagestyle{fancy}
\rhead{\includegraphics[width=2.5cm]{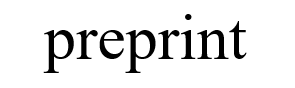}}

\title{Population fluctuation mechanism of  the super-thermal photon statistic of quantum LEDs with collective effects}

\maketitle

\author{Igor E. Protsenko*}
\author{Alexander V. Uskov}

\begin{affiliations}
 P.N.Lebedev Physical Institute of the RAS,
Moscow 119991, Russia\\
Email Address: procenkoie@lebedev.ru, protsenk@gmail.com

\end{affiliations}


\begin{abstract}

The analytical procedure has been developed for the calculation of the quantum second-order autocorrelation function $g_2$ of a small super-radiant LED, where the field, polarisation, and population of the active medium cannot be eliminated adiabatically. The Langevin force, which describes the effect of population fluctuations on the LED polarisation and preserves the operator commutation relations, has been found. It is demonstrated that the super-thermal photon statistics with $g_2 > 2$ of a small quantum LED with large photon number fluctuations, emitter-field coupling, bad cavity, and operating in the limit $n \rightarrow 0$, is the result of a combined effect of the spontaneous emission, collective effects and population fluctuations. Analytical expressions for $g_2$ and $n$ are derived.

\end{abstract}
\section{\label{Sec1}Introduction}
The second-order correlation function $g_2$ is an essential characteristic of quantum light  \ct{Huang:16,RevModPhys.70.101}. 
Measurements of $g_2$ play an important role in emerging research in quantum information \ct{PhysRevLett.97.083604}, quantum dot fluorescence correlation spectroscopy \ct{Michler2000}, cold atomic clouds \ct{Das:10}, single molecules \ct{PhysRevLett.76.900}, optical sensing \ct{9466840} and nano-lasers \ct{George:21,Hayenga:16}.
In the quantum theory $g_2=\left<\hat{a}^+\hat{a}^+\hat{a}\hat{a}\right>/n^2$. Here and below $\hat{a}$ is the Bose operator of the electromagnetic field, $n=\left<\hat{a}^+\hat{a}\right>$ is the mean number of photons, the notation $\left<...\right>$ means quantum mechanical averaging. Here we are looking at $g_2$ with a zero delay, i.e. when the operators in $g_2$ are taken at the same moment in time.
It is known that $g_2=1$ for coherent, $g_2=2$ for thermal and $g_2>2$ for super-thermal light \ct{Hertel,Fox2006}. The super-thermal photon statistics show the collective effects as super- or sub-radiance \ct{Jahnke}. The study of the mechanisms of super-thermal photon statistics helps to understand the physics of quantum emitters with collective effects and to improve their performance.

There have been several papers on the theoretical modeling and experimental measurements of how various aspects of laser (beyond spontaneous emission): pump, loss, and population 
fluctuations (as multiplicative noise), cooperative atomic effects, etc., affect $g_2$ and cause it
to be super-thermal, going back to the early days of the laser theory and continuing through the 80s and 90s. 
For example, the superthermal photon statistics have been  observed experimentally in a dye laser \ct{PhysRevLett.52.341} and  described theoretically in \ct{PhysRevLett.52.341, PhysRevA.35.1838}. 
The model for $g_2$ calculations in \ct{PhysRevLett.52.341, PhysRevA.35.1838} is  limited: the laser field  is only a  dynamic variable. 
The  presentation and discussion of all previous work on $g_2$ and the field intensity fluctuations in lasers is a broad subject and  requires a review paper. 
For example, calculations of the quantum $g_2$ for the superradiant emitters and high-$\beta$ nano-lasers are presented in \ct{Auffeves_2011,PhysRevA.81.063827,PhysRevApplied.10.054055,PhysRevA.99.053820,Kreinberg2017}.

Recently, there has been interest in a small bad-cavity laser where  the superradiance (SR) is significant \ct{Jahnke,PhysRevX.6.011025,PhysRevA.96.013847,PhysRevA.81.033847,PhysRevA.98.063837,Bohnet}. Such SR-lasers exhibit the superthermal photon statistics with $g_2>2$ below the threshold in a  quantum regime, where  $n\rightarrow 0$ and quantum fluctuations are large \ct{Kreinberg2017,Jahnke, Bhatti2015}. 

Measurements of the ultrafast second-order
correlations of broadband amplified spontaneous emission from a superluminescent
diode 
revealed an exciting hybrid coherent photon state with reduced
photon bunching \ct{Blazek:11}. The power-dependent second-order
coherence for two metallic nanolasers close to the thresholdless lasing regime has been studied in \ct{Kreinberg_01}, where the power dependence of the $\beta-$factor due to phase space filling effects was found. It is worth mentioning a very recent paper \ct{PhysRevLett.130.253801} which used a stochastic approach to the quantum noise of a single-emitter nanolaser to find $g_2$.

$g_2$ calculations for small lasers and LEDs are more difficult than in early work, e.q. \ct{PhysRevLett.52.341, PhysRevA.35.1838}, which dealt with macroscopic lasers. The quantum model of a small bad-cavity SR laser includes  the field, the polarisation and the population as dynamical variables -- operators whose  commutation relations must be preserved.  
It is a non-trivial and not only a technical task to include all dynamical variables without adiabatic elimination in the quantum laser model, as the model of \ct{PhysRev.185.568}, especially with the analytical treatment of higher order correlations, as in $g_2$.
An example of a  $g_2$ calculation method is  the cluster (or cumulant) expansion approach   \ct{PhysRevApplied.4.044018, 10.1063/1.5138937,Leymann_01}, which is suitable for numerical calculations \ct{Plankensteiner2022quantumcumulantsjl}.
The application of the cluster expansion method in our model hardly leads to a compact analytical result for $g_2$ due to the large number of equations.  
In Section 2 we explain some other difficulties in the computation of $g_2$ that are encountered in our approach. We overcome some of them in the present paper.

A model suitable  for analytical calculations of quantum $g_2$ for LEDs and lasers with the field, polarisation and  population dynamics  is  not yet fully developed. 
The aim of this paper is to make the theory of superradiant lasers \ct{PhysRevA.59.1667,Andre:19,Protsenko_2021,PhysRevA.105.053713,https://doi.org/10.1002/andp.202200298}, with all dynamical variables, suitable for the calculation of $g_2$ and other high order correlations, including the super-thermal $g_2>2$ in the quantum regime, when $n\rightarrow 0$. 
The approach must preserve the commutation relations and lead, in some approximations, to analytical results. 
Then, it is a contribution to the understanding of the conditions and physical mechanisms of super-thermal photon statistics, in particular the role of population fluctuations and collective effects.  

We assume a large number of emitters and neglect correlations between the population and the polarisation fluctuations, which simplifies the calculations.  
This differs from the superradiant laser models where only one or a few emitters are placed in the cavity as shown in \ct{PhysRevA.18.1628} and in the recent paper \ct{PhysRevLett.130.253801}.

This paper investigates fluctuations in the number of emitters as the physical mechanism of superthermal photon statistics.
The emitter number fluctuations in our model are due to the  incoherent excitation of emitters from their lower to their upper states and the decay of the excited  states, for example by  spontaneous emission.

We find that the increase in the mean photon number $n$ due to the population fluctuations \ct{PhysRevA.105.053713} is closely related to the superthermal photon statistics. 
Analytical formulae are derived for $n$ and $g_2$. 
We study a small LED in the linear regime, i.e. when the LED pump is weak and the population fluctuations  do not depend on the field in the LED cavity. 
It is well known that at such low excitation super thermal photon statistics appear   \ct{PhysRevA.101.063835}. 
The approach presented below can be easily generalised to high excitation in the nonlinear laser regime.

In Section~\ref{Sec2}, we present the basic equations, explain the notations, review our previous results, and explain the difficulties  in calculating $g_2$ that prevent us from finding the correct $g_2$ in previous work. 
We describe three tasks completed in this paper for the calculation of $g_2$ and present a new explicit expression for the polarisation fluctuation Langevin force with an account of the population fluctuations. 
See Appendix~\ref{LF_FN} for the mathematical details of Section~\ref{Sec2}.

In section~\ref{Sec3} we derive a new  analytical formula \rf{mean_n_norm} for the  increase $\Delta_n$ of the mean  photon number caused by the population fluctuations.
We see that a large $\Delta_n$ appears under conditions favourable for collective effects \ct{Belyanin_1998,Koch_2017,Khanin}. 
The parameter $\Delta_n$ is shown in Figure~\ref{Fig_3} as a function of the "adiabaticity parameter"  $2\kappa/\gamma_{\perp}$, which is large for strong collective effects. 
The expression~\rf{mean_n_norm} for $\Delta_n$ will be used for the calculation of $g_2$ in section~\ref{Sec4}.

The second order autocorrelation function $g_2$, given by Eq.~\rf{g2_fin}, is found in   Section~\ref{Sec4}, with the main mathematical part of the calculations presented in  Appendix~\ref{g2_der}. Examples of the super-thermal $g_2$ at various LED parameters are shown in Figs~\ref{Fig_4} and \ref{Fig_5}.

The approach and the  results  are discussed in Section~\ref{Section_5}, and the conclusion  completes the paper. The main conclusion is that, in contrast to previous work, the super-thermal photon statistics ($g_2>2$) in LEDs  operating in the quantum regime ($n\ll 1$) is a consequence of the population fluctuations  together with the combined effect of non-adiabatic dynamics of  polarisation and the collective effects in the radiation  of the emitters. 
\section{\label{Sec2} Basic equations, approximations  and the  polarisation Langevin force}
\subsection{Basic equations}
We consider the model of a stationary homogeneously broadened single-mode laser with $N_0\gg 1$ identical two-level emitters.

Lasing transitions  are resonant with the cavity mode of the optical carrier frequency $\omega_0$.  $\hat{a}(t){{e}^{-i{{\omega }_{0}}t}}$   is the
Bose-operator of the cavity mode with amplitude $\hat{a}(t)$ changes much slower than ${{e}^{-i{{\omega }_{0}}t}}$.  We describe the laser by the theory of \ct{PhysRevA.59.1667, Andre:19,Protsenko_2021,PhysRevA.105.053713} to be  similar to the original model of the laser
noise as in  \ct{PhysRev.185.568} and other papers
in this series. In contrast to \ct{PhysRev.185.568}, we consider the polarisation of the active medium as a dynamical variable, which is necessary for the analysis of small bad-cavity superradiant lasers.

The laser equations are
\begin{subequations}\lb{MBE_0}\beqr
  \dot{\hat{a}}&=&-\kappa \hat{a}+\Omega \hat{v}+\sqrt{2\kappa}\hat{a}_{in} \lb{MBE_1}\\
 \dot{\hat{v}}&=&-\left( {{\gamma }_{\bot }}/2 \right)\hat{v}+\Omega f\hat{a}\left(N+2{{\delta{\hat{N}}}_{e}} \right)+{{{\hat{F}}}_{v}} \lb{MBE_2}\\ 
 \delta{{{\dot{\hat{N}}}}_{e}}&=&-\Omega \delta\hat{\Sigma}-{{\gamma }_P{{\delta{\hat{N}}}_{e}}} +{{{\hat{F}}}_{N_e}}, \lb{MBE_3} 
\eeqr\end{subequations}
where $\hat{v}$ and $\delta\hat{N}_e$ are the polarisation of the active medium and the population fluctuation operators of the emitter excited states, respectively;  $\Omega$ is the vacuum Rabi frequency,  $\hat{a}_{in}$ is the  Bose operator of the vacuum field entering the cavity through the semitransparent mirror; $\kappa$, ${\gamma }_{\bot }$ and  ${\gamma }_{\parallel }$ are the cavity mode, the polarisation, and the population decay rates, respectively, 
\beq
{\gamma }_P = {\gamma }_{\parallel }(P+1), \lb{gamma_P}   
\eeq
$P$ is the dimensionless pumping rate, $N = N_e-N_g$ is the mean population inversion, $N_e$ ($N_g$) are the mean populations of the upper (lower) emitter states, $N_e+N_g=N_0$;  ${{\hat{F}}_{\alpha }}$ with  $\alpha =\left\{v,{{N}_{e}} \right\}$  are the Langevin force  operators;    $\delta\hat{\Sigma}$ denotes  fluctuations of the operator $\hat{\Sigma} = {{\hat{a}}^{+}}\hat{v}+{{\hat{v}}^{+}}\hat{a}$,  $f\approx 1/2$. The derivation of Eqs.~\rf{MBE_0} is given in \ct{Andre:19,Protsenko_2021,PhysRevA.105.053713}.

The law of conservation of energy follows from Eqs.~\rf{MBE_0}
\beq
        2\kappa n=  \gamma_{\parallel}[P(N_0-N_e)-N_e], \lb{eql_1}
\eeq
where the mean cavity photon number  $n=\left<\hat{a}^+\hat{a}\right>$. 
The operator $\hat{a}$ can be found by using Eqs.~\rf{MBE_0} as a function of $N_e$, and then $N_e$ and $n$ can be determined from Eq.~\rf{eql_1}.

The stationary solution of equations \rf{MBE_0} and \rf{eql_1} can be found in a variety of approximations. For example, in semi-classical laser theory, the operators $\hat{a}$ and $\hat{v}$ can be replaced by c-numbers, thereby neglecting quantum fluctuations \ct{Scully,Yariv}.
It is possible to consider quantum fluctuations at levels well above the lasing threshold, provided that the equations~\rf{MBE_0} are linearised around the steady-state solution of the semi-classical laser theory  \ct{1071726, RevModPhys.68.127, PhysRevA.47.1431}.
In the case where $2\kappa \ll \gamma_{\perp}$ and the active medium polarisation can be eliminated from Eqs.~\rf{MBE_0} adiabatically, a rate equation approach can be employed  \ct{PhysRevLett.130.253801,Coldren,10.1063/1.5022958,atoms9010006}. 

The aim of the present study is to preserve quantum fluctuations and polarisation dynamics, to consider a weak radiation below the laser threshold, and to solve Eqs.~\rf{MBE_0} by a perturbation approach on population fluctuations. 
The following subsection presents a "zero-order" approximation on the population fluctuations and elucidates the reasons why this and some other approximations are inadequate for $g_2$ calculations in the quantum regime when the mean photon number is small.  
\subsection{"Zero order" and other approximations}
In a "zero order" approximation, we neglect population fluctuations. When  $\delta\hat{N}_e = 0$, Eqs.~\rf{MBE_1}, \rf{MBE_2} are linear in $\hat{a}$, $\hat{v}$ operators and solved by the  Fourier transform \ct{PhysRevA.59.1667, Andre:19,Protsenko_2021}. For $\delta\hat{N}_e = 0$ polarisation  Langevin force is  
\beq
\hat{F}_v =\hat{F}_v^{(0)} = \sqrt{2\gamma_{\perp}fN_e}\hat{b}^{(0)+}_{in}. \lb{F_v0}
\eeq
Fourier components  $\hat{b}_{in}^{(0)}(\omega)$, ($(\hat{b}_{in}^{(0)+})_{\omega}$) of operators $\hat{b}_{in}^{(0)}$ ($\hat{b}_{in}^{(0)+}$) obey a free space Bose commutation relations $[\hat{b}_{in}^{(0)}(\omega),(\hat{b}_{in}^{(0)+})_{\omega'}] = \delta(\omega+\omega')$. 
$\hat{b}_{in}^{(0)}$ is the amplitude  Bose operator of the vacuum  bath of "inverted" oscillators. The Bose operator of  the inverted oscillator is $\hat{b}_{in}^{(0)}e^{i\omega_0t}$. Note the $"+"$ sign  in $e^{i\omega_0t}$ for the inverted oscillator, while it is $e^{-i\omega_0t}$ for the normal oscillator.  An inverted oscillator bath has been introduced in \ct{Glauber}. It has been used  in the theory of  quantum amplifiers \ct{Stenholm_1986} and lasers \ct{https://doi.org/10.1002/andp.202200298}. The expression Eq.~\rf{F_v0}  is justified in \ct{https://doi.org/10.1002/andp.202200298}. 

A "zero order" approximation  leads to interesting results. It finds  $n$, the photon number spectrum $n(\omega)$ analytically for the thresholdless laser  with the polarisation to be a dynamical variable \ct{PhysRevA.59.1667} and predicts a new effect of the collective Rabi splitting in the superradiant LEDs \ct{Andre:19}. However, this approximation  does not  properly describe  high-order correlations and, in particular,  leads to  $g_2 = 2$ for all values of the laser parameters. This can be found by calculating $\hat{a}$  and  $\left<\hat{a}^+\hat{a}^+\hat{a}\hat{a}\right>$ from Eqs.~\rf{MBE_0} with  $\delta\hat{N}_e = 0$. Meanwhile, superradiant lasers show superthermal photon statistics in experiments with $g_2>2$ as $n\rightarrow 0$ \ct{Kreinberg2017,Jahnke, Bhatti2015}.

In this paper, we bring our theory into a qualitative agreement with experimental results as \ct{Kreinberg2017,Jahnke, Bhatti2015}, i.e. we find $g_2>2$ as  $n\rightarrow 0$ under some conditions, and explain why this happens in the framework of our model. Let us briefly review some other our previous findings to show the place of the present work in the study.

Since we have calculated $g_2=2$ without the population fluctuations, we must include the population fluctuations  to find the super-thermal photon statistics in the theory. We know how to add the population fluctuations above the laser threshold to the theory when Eqs.~\rf{MBE_0} can be linearized around the steady state mean values. We can then proceed with a small signal analysis, well known in laser theory and quantum optics   \ct{1071726, RevModPhys.68.127, PhysRevA.47.1431}. In \ct{Protsenko_2021} we consider the linearized quantum laser equations with the field, polarisation, and population dynamics above the threshold,  and find an interesting result, that a well-known $1/2$ difference in the laser linewidth expressions below and above the threshold  \ct{Yariv} is related to the population fluctuations. 
In \ct{PhysRevA.105.053713} we conclude, that the small-signal analysis with the population fluctuations   used in \ct{Protsenko_2021} cannot be performed correctly  below the threshold, where the field and polarisation are dynamical variables with zero mean. 

In \ct{PhysRevA.105.053713} we return to the nonlinear quantum laser equations \rf{MBE_0} and find the polarisation Langevin force power spectra (or diffusion coefficients) for $\hat{F}_v$ in Eq.~\rf{MBE_2} with the population fluctuations. We did this using the requirement that the Bose commutation relations for $\hat{a}$ must be preserved, when the population fluctuations are taken into account. New diffusion coefficients allow a new finding that population fluctuations significantly increase the emitted field in the superradiant LED under certain conditions.   However, the diffusion coefficients found in \ct{PhysRevA.105.053713} are not sufficient for the calculation of higher order correlations as in $g_2\sim\left<\hat{a}^+\hat{a}^+\hat{a}\hat{a}\right>$. This can be seen by following the procedure of the $g_2$ calculations in the Appendix \ref{g2_der} which, in fact, does not include the $\hat{F}_v$ diffusion coefficients. By examining the calculations of the cumulants   in the  in Appendix~\ref{g2_der} (see Eqs.~\rf{cumulant_calcul}), we suggest that $\hat{F}_v$, with the population fluctuations, corresponds to a non-Gaussian noise. Therefore, the knowing of the power spectra (the first order correlations or the diffusion coefficients) of $\hat{F}_v$ is  not sufficient to calculate of the second- and higher order correlations. We note, that the correct calculation of $g_2$ is not possible using only the equations for the photon number fluctuations $\delta n$ presented in \ct{PhysRevA.105.053713}. Such equations correspond to a zero-order approximation, lead to $g_2=2$, their purpose is to find $\delta 
\hat{N}_e$ by the perturbation approach. Then $\delta 
\hat{N}_e$ can be used to find $g_2$ by the procedure described in this paper. This procedure requires an explicit expression for  $\hat{F}_v$,  which is given below. 

In summary, this paper completes three tasks not presented in our previous work.  The first task is to find an explicit expression for $\hat{F}_v$ with the population fluctuations. The second task is to perform  $g_2$ calculations  showing  the super-thermal photon statistics ($g_2>2$)  and the role of the population fluctuations in the super-thermal statistics in the quantum ($n\rightarrow 0$) regime. The third task is  to derive the analytical expression for $g_2$, in some approximations, which explicitly gives conditions for the super-thermal photon statistics as $n\rightarrow 0$. 

In this paper we restrict ourselves to  $n\ll 1$ and do not consider the nonlinear radiation regime. Therefore, we neglect the stimulated emission, drop the term $\sim \hat{a}\delta \hat{N}_e$ in Eq.~\rf{MBE_2}, but use the Langevin force $\hat{F}_v$, which depends on the population fluctuations. In this way  we preserve the population fluctuation effect on the  {\em spontaneous emission} to the cavity mode, which is the main source of radiation as $n\rightarrow 0$. 
\subsection{Polarisation fluctuation Langevin force}
Let us find the  explicit expression for  the polarisation Langevin force $\hat{F}_v$ with the population fluctuations. We represent $\hat{F}_v$   as a sum 
\beq
\hat{F}_v= \hat{F}_v^{(0)}+\hat{F}_v^{(N)}, \lb{f_v_sum}
\eeq
where $\hat{F}_v^{(N)}$ describes the polarisation noise associated with the  population fluctuations, and $\hat{F}_v^{(0)}$ is given by Eq.~\rf{F_v0}. We assume a large number of emitters $N_0\gg 1$, each emitter interacting with its own bath uncorrelated with the baths of other emitters. We therefore neglect correlations between $\hat{F}_v^{(0)}$ and $\hat{F}_v^{(N)}$ with good accuracy $\sim N_0^{-1}$. With the same precision,  we neglect the correlations between the population and the polarisation fluctuations.

We propose the structure of $\hat{F}_v^{(N)}$ by the analogy with the structure of   $\hat{F}_v^{(0)}$ given by Eq.~\rf{F_v0}. We see that $\hat{F}_v^{(0)}\sim \hat{b}^{(0)+}_{in}$, where  $\hat{b}^{(0)}_{in}$ is a  Bose-operator of the inverted oscillator associated with the vacuum fluctuations of the polarisation dephasing noise. By the analogy with $\hat{F}_v^{(0)}$ we assume that $\hat{F}_v^{(N)} \sim\hat{b}^+_{in}$, where $\hat{b}_{in}$ is the vacuum Bose operator of the inverted oscillator, related to the polarisation noise caused by the population fluctuations. The assumption that $\hat{F}_v^{(N)} \sim\hat{b}^+_{in}$ and the requirement of Bose commutation relations for $\hat{a}$ lead to $\hat{F}_v^{(N)}$, which is found in Appendix~\ref{LF_FN}
\beq
\hat{F}_v^{(N)} = 2\Omega f \hat{b}_{in}^+\delta \hat{N}_e. \lb{f_v_expr}
\eeq
The power spectra of $\hat{b}_{in}$  are
\beq
\left< (\hat{b}_{in}^+)_{\omega}\hat{b}_{in}(\omega')\right>=0, \hspace{0.5cm} \left<\hat{b}_{in}(\omega)(\hat{b}_{in}^+)_{\omega'}\right>=c(\omega)\delta(\omega+\omega'), 
\lb{b_in_PS}
\eeq
where $\hat{b}_{in}(\omega)$ and $(\hat{b}_{in}^+)_{\omega}=\hat{b}_{in}^+(-\omega)$ are Fourier components of the $\hat{b}_{in}(t)$ and $\hat{b}_{in}^+(t)$ operators  respectively;
\beq 
c(\omega )=\frac{2\kappa {{\omega }^{2}}+(\kappa \gamma _{\bot }^{2}/2)(1-N/N_{th}) }{{{\left| s(\omega ) \right|}^{2}}}, \lb{comm_sp}
\eeq
and $s(\omega )$ is given by Eq.~\rf{s_small}. We see that $(2\pi)^{-1}\int_{-\infty}^{\infty}c(\omega)d\omega = 1$, so 
the Bose commutation relations    
\[\left<[\hat{b}_{in}(t), \hat{b}^+_{in}(t)]\right>=1\]
are satisfied. It is shown in Appendix~\ref{LF_FN} that the Langevin force $\hat{F}_v$, the  Bose commutation relations  for $\hat{a}$ and the commutation relations $\left<[\hat{v},\hat{v}^+]\right> = f(N_g-N_e)$  found in \ct{https://doi.org/10.1002/andp.202200298} are preserved.

Eqs.~\rf{F_v0} -- \rf{comm_sp} give the explicit expression for the polarisation Langevin force $\hat{F}_v$ taking into account the population fluctuations. These equations and the correlation properties of  $\delta \hat{N}_e$  allow to find higher order correlations of operators in Eqs.~\rf{MBE_0}, in particular  $\left<\hat{a}^+\hat{a}^+\hat{a}\hat{a}\right>$ and $g_2$.  
\section{\label{Sec3}  The photon number spectrum and the mean photon number}
In \ct{PhysRevA.105.053713} it was found  that the population fluctuations increase the cavity mode spontaneous emission and the mean photon number $n$. Here we  derive analytical expressions (not presented in \ct{PhysRevA.105.053713}) for the field spectrum $n(\omega)$ and $n$  below the threshold with the population fluctuations. These expressions are needed for the calculation of $g_2$ in Section \ref{Sec4}.  

Substituting the expression \rf{a_FC} for the field amplitude Fourier component operator $\hat{a}(\omega)$ from Eq.~\rf{a_FC}  into  
\beq
\left<(\hat{a}^+)_{\omega}\hat{a}(\omega')\right>=n(\omega,N_e)\delta(\omega+\omega'), \lb{n_omega_g_expr}
\eeq
we obtain the integral equation for $n(\omega,N_e)$ with the  unknown population fluctuation spectrum $\delta^2N_e(\omega)$ and $N_e$. $\delta^2N_e(\omega)$ can be found by a perturbation procedure described in \ct{PhysRevA.105.053713}. Then we calculate $n(\omega,N_e)$, $n(N_e)=(2\pi)^{-1}\int_{-\infty}^{\infty}n(\omega,N_e)d\omega$, find $N_e$ and  $n$  from the energy conservation law \rf{eql_1}. 

We consider a weak pump and a linear radiation regime (the LED regime), where the number $n$ of cavity photons $n\rightarrow 0$. Thus we  neglect the nonlinear terms proportional to  $\hat{a}$  and $\hat{a}(\omega)$ in Eqs.~\rf{FC_2}, \rf{a_FC}. Such non-linear terms are usually neglected in the analysis of amplifiers \ct{Stenholm_1986}. We retain the population fluctuation dependent term $\hat{F}^{(N)}_v$ in the Langevin force \rf{f_v_sum}. Physically, we neglect  the stimulated but preserve the spontaneous emission (including its population-dependent part) to the cavity mode. For simplicity, we remove  the term with the input vacuum fluctuation Langevin force $\hat{F}_a$ from Eq.~\rf{a_FC}, since  it only appears  in the combinations  $\hat{F}_{a^+}\hat{F}_a$, $\left<\hat{F}_{a^+}\hat{F}_a\right>=0$, so such a term does not contribute to the final results. So we take
\beq
\hat{a}(\omega )=\frac{\Omega \hat{F}_v^{(0)}(\omega )+(\kappa\gamma_{\bot }/N_{th})\left(\hat{b}_{in}^+\delta\hat{N}_e \right)_{\omega }}{s(\omega )}, \lb{a_FC_linear}
\eeq
where $s(\omega)$ is given by Eq.~\rf{s_small}. In the linear regime, we drop $\Omega \delta\hat{\Sigma }(\omega)$ in Eq.~\rf{FC_3} and find

$\delta\hat{N}_e(\omega)=\hat{F}_{N_e}(\omega)/(i\omega-\gamma_P)$.
The power spectrum of the population fluctuation Langevin force is
\beq
2D_{N_eN_e} = \gamma_{\parallel}(PN_g+N_e), \lb{2DNN}
\eeq
which is the same as in the rate equation laser theory \ct{Coldren}. Eq.~\rf{2DNN} is a good approximation for $2D_{N_eN_e}$ with a large number of emitters. The power spectrum of the population fluctuations is
\beq
    \delta^2{N}_e(\omega) = 2\gamma_P\delta^2{N}_e/(\omega^2+\gamma_P^2), \lb{p_fp_p_sp}
\eeq
where $\gamma_P$ is given by Eq.~\rf{gamma_P}. Using Eq.~\rf{2DNN} we find the upper state population fluctuation dispersion $\delta^2{N}_e$ and the mean population $N_e$ of the upper emitter states 
\beq
\delta^2{N}_e = N_e/(P+1), \hspace{0.5cm} N_e = PN_0/(P+1). \lb{pop_fl_params}
\eeq
We substitute $\hat{a}(\omega )$ from Eq.~\rf{a_FC_linear} and the Fourier component $(\hat{a}^+)_{\omega}\equiv\hat {a}^+(-\omega )$   in Eq.~\rf{n_omega_g_expr} for $n(\omega)$, neglect by correlations between $\hat{F}_v^{(0)}(\omega )$, $\left(\hat{b}_{in}^+\delta\hat{N}_e \right)_{\omega }$  and find $n(\omega)$. The  power spectrum $S_{bN_e}(\omega)$ of the operator product  $\hat{b}_v\delta\hat{N}_e$  is given by
\beq
S_{bN_e}(\omega)\equiv(c*\delta^2 N_e)_{\omega} =\frac{1}{2\pi}\int_{-\infty}^{\infty}c(\omega-\omega')\delta^2 N_e(\omega')d\omega'. \lb{conv_00}
\eeq
We simplify the integral \rf{conv_00} by noting  that the width of the $c(\omega)$ spectrum  is  $\sim\sqrt{\kappa\gamma_{\perp}} \gg \gamma_P\approx\gamma_{\parallel}$, which is the width of the population fluctuation spectrum  $\delta^2N_e(\omega)$. So we approximate
\beq
(c*\delta^2 N_e)_{\omega}\approx c(\omega)\delta^2N_e \lb{approx_conv} 
\eeq
under the condition 
\beq
\gamma_{\parallel}/\sqrt{\kappa\gamma_{\perp}}\ll 1. \lb{gamma_par_cond}
\eeq
Substituting $\hat{a}(\omega)$ from Eq.~\rf{a_FC_linear} into $\left<(\hat{a}^+)_{\omega}\hat{a}(\omega')\right>=n(\omega)\delta{(\omega+\omega')}$ we find the photon number spectrum 
\beq
n(\omega )=\frac{\left( \kappa \gamma _{\bot }^{2}/2{{N}_{th}} \right){{N}_{e}}+\delta^2N_e{{(\kappa {{\gamma }_{\bot }}/{{N}_{th}})}^{2}}c(\omega)}{{{\left| s(\omega ) \right|}^{2}}} \lb{n_omega}
\eeq
and calculate the mean photon number
\beq
n=\frac{1}{2\pi}\int_{-\infty}^{\infty}n(\omega)d\omega = n_0 + \delta^2N_e\left(\frac{\kappa\gamma_{\perp}}{N_{th}}\right)^2\frac{1}{2\pi}\int_{-\infty}^{\infty}\frac{c(\omega)}{{{\left| s(\omega ) \right|}^{2}}}d\omega. \lb{n_pop_fl_0}
\eeq
We compute the integral in Eq.~\rf{n_pop_fl_0}, perform some algebraic transformations 
and find 
\beq
n=n_0\left(1+\Delta_n \right), \hspace{0.5cm}  \Delta_n=\frac{\delta^2N_e}{N_e}\frac{1}{N_{th}}\left[3\left(\frac{2\kappa/\gamma_{\perp}}{1+2\kappa/\gamma_{\perp}}\right)^2 + \frac{2\kappa/\gamma_{\perp}}{1-
N/N_{th}}\right]\lb{mean_n_norm}
\eeq
where $\Delta_n$ is a relative increase of $n$  caused by  population fluctuations and 
\beq
n_0=\frac{{{N}_{e}}}{\left( 1+2\kappa /{{\gamma }_{\bot }} \right)\left( {{N}_{th}}-N \right)} \lb{mean_n_0}
\eeq
is a part of the mean photon number found in the "zero-order" approximation, i.e. by  Eqs.~\rf{MBE_0} with $\delta\hat{N}_e=0$; $n_0$ is the same as in \ct{Andre:19,PhysRevA.105.053713,Protsenko_2021}.  According to  Eqs.~\rf{pop_fl_params} ${\delta^2N_e}/{N_e} = 1/(P+1)\approx 1$ for a weak pump $P\ll 1$.

\begin{figure}[thb]\bc
\includegraphics[width=9cm]{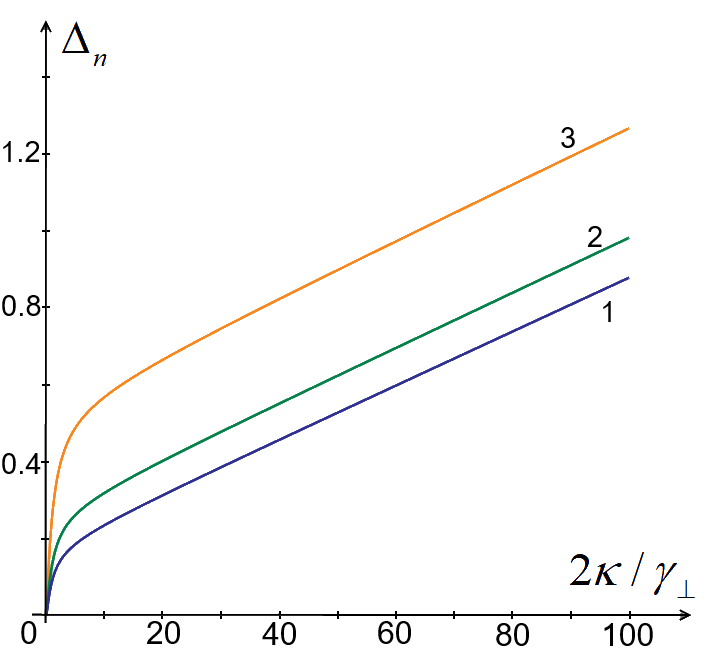}
\caption{The relative increase $\Delta_n$ of the cavity photon number  due to the population fluctuations  at $P=0.1$, $\gamma_{\parallel}/\gamma_{\perp} = 0.1$ as a function of  $2\kappa/\gamma_{\perp}$ at   $N_{th}=15$ (curve 1), $10$ (2) and $5$ (3).       }
\label{Fig_3}\ec
\end{figure}

The expression \rf{mean_n_norm}  shows that the population fluctuations increase significantly $n$ for a small $N_{th}$ and for a bad cavity when $2\kappa/\gamma_{\perp}$ is large, $2\kappa/\gamma_{\perp}>1$, which is typical for the super-radiant LEDs and lasers. A small $N_{th}$ implies a large field-emitter coupling constant $\Omega \sim N_{th}^{-1/2}$.   Fig.~\ref{Fig_3} shows $\Delta_n$ as a function of $2\kappa/\gamma_{\perp}$ for different  $N_{th}$. 
%
%
%
%
\section{\label{Sec4} Super-thermal photon statistics}
We  use the expression \rf{a_FC_linear}  to calculate the second order autocorrelation function \newline
$g_2=\left<\hat{a}^+(t)\hat{a}^+(t)\hat{a}(t)\hat{a}(t)\right>/n^2$. Doing the Fourier expansion of $\hat{a}(t)$ we write
\beq
g_2 = \frac{1}{n^2}\frac{1}{(2\pi)^2}\int_{-\infty}^{\infty}\left<(\hat{a}^+)_{\omega_1}(\hat{a}^+)_{\omega_2}\hat{a}(\omega_3)\hat{a}(\omega_4)\right>e^{-i(\omega_1+\omega_2+\omega_3+\omega_4)t}d\omega_1d\omega_2d\omega_3d\omega_4. \lb{g2_def1}
\eeq
The stationary $g_2$ does not depend on time, so it must be $\omega_1+\omega_2+\omega_3+\omega_4=0$ in Eq.~\rf{g2_def1}. Therefore,  we drop the exponent multiplier in \rf{g2_def1} and write
\beq
g_2 = \frac{1}{n^2}\frac{1}{(2\pi)^2}\int_{-\infty}^{\infty}\left<(\hat{a}^+)_{\omega_1}(\hat{a}^+)_{\omega_2}\hat{a}(\omega_3)\hat{a}(\omega_4)\right>_{\omega_1+\omega_2+\omega_3+\omega_4=0}d\omega_1d\omega_2d\omega_3d\omega_4. \lb{g2_def2}
\eeq
For the sake of simplicity, we drop the index $\omega_1+\omega_2+\omega_3+\omega_4=0$ in Eq.~\rf{g2_def2} and in all integrals in the calculations of $g_2$ but remember the condition represented by this index. We find $\left<(\hat{a}^+)_{\omega_1}(\hat{a}^+)_{\omega_2}\hat{a}(\omega_3)\hat{a}(\omega_4)\right>$, calculate the integral in Eq.~\rf{g2_def2} and  obtain $g_2$.

Substituting the expression \rf{a_FC_linear} for $\hat{a}(\omega)$ into Eq.~\rf{g2_def2} and neglecting the correlations between $\hat{F}_v^{(0)}(\omega )$ and $\left(\hat{b}_{in}^+\delta\hat{N}_e \right)_{\omega }$,  we find
\beq
g_2 = 2 + \left(\frac{\kappa\gamma_{\perp}}{N_{th}}\right)^4\frac{C_g}{n^2}, \lb{g_2_cummul}
\eeq
where $C_g$ is a cumulant for the operator $\hat{B}(\omega) = \left(\hat{b}_{in}^+\delta\hat{N}_e\right)_{\omega}/s(\omega)$:
\[
C_g = \frac{1}{(2\pi)^2}\int_{-\infty}^{\infty}\left<(\hat{B}^+)_{\omega_1}(\hat{B}^+)_{\omega_2}\hat{B}(\omega_3)\hat{B}(\omega_4)\right>d\omega_1...d\omega_4\]\beq - 2\left[\frac{1}{2\pi}\int_{-\infty}^{\infty}\left<(\hat{B}^+)_{\omega_1}\hat{B}(\omega_2)\right>d\omega_1d\omega_2\right]^2. \lb{cumulant_g}
\eeq
Here 
\[
\left<(\hat{B}^+)_{\omega_1}\hat{B}(\omega_2)\right> = \frac{S_{bN_e}(\omega)}{|s(\omega)|^2}\delta(\omega_1+\omega_2)
\]
and the spectrum $S_{bN_e}(\omega)$ of the operator product $\hat{b}_{in}^+\delta\hat{N}_e$ is  determined by Eq.~\rf{conv_00}. Calculations of $C_g$ and some algebraic transformations shown in Appendix~\ref{g2_der} lead to 
\beq
g_2 = 2\left[1+2\left(\frac{\Delta_n}{1+\Delta_n}\right)^2\right], \lb{g2_fin}
\eeq
where  $\Delta_n$  is given by Eq.~\rf{mean_n_norm}. The result \rf{g2_fin} is found in the approximation \rf{approx_conv} valid under the condition \rf{gamma_par_cond}. 

\begin{figure}[thb]\bc
\includegraphics[width=9cm]{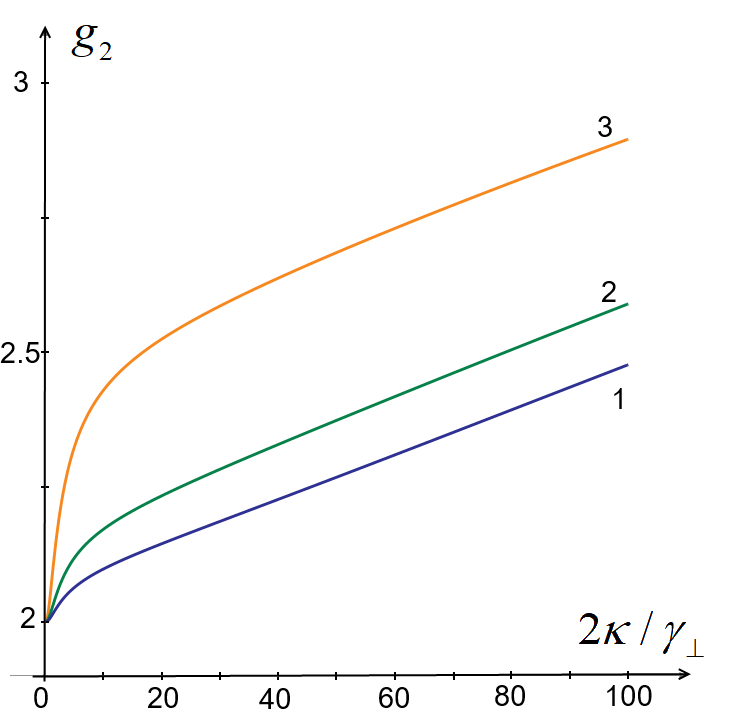}
\caption{$g_2>2$ at a weak pump $P=0.1$ as a function of  $2\kappa/\gamma_{\perp}$ for  $N_{th} = 15$ (curve 1), $10$ (curve 2) and $5$ (curve 3).       }
\label{Fig_4}\ec
\end{figure}

\begin{figure}[thb]\bc
\includegraphics[width=9cm]{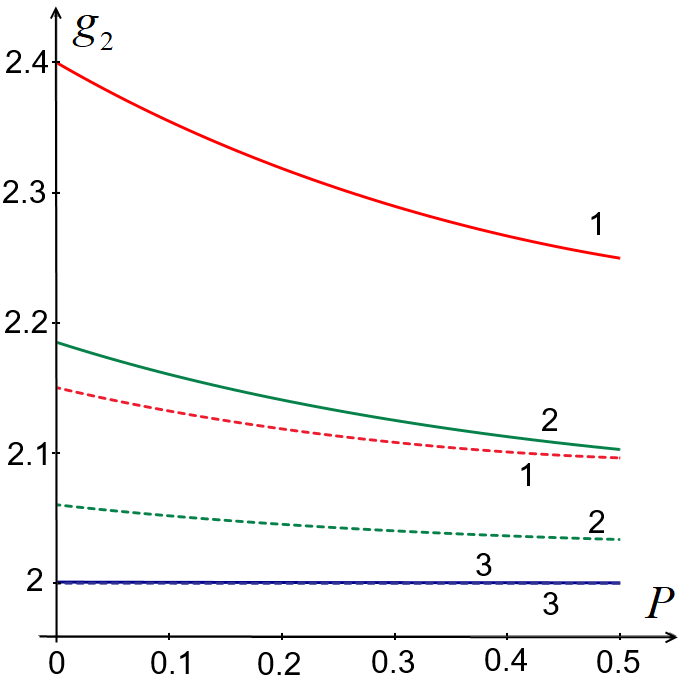}
\caption{$g_2$ as a function of  $2\kappa/\gamma_{\perp} = 6$ (curves 1), $2$ (curves 2) and $0.2$  (curves 3). $N_{th} = 5$ for the solid curves and $10$ for the dashed curves.       }
\label{Fig_5}\ec
\end{figure}

Fig.~\ref{Fig_4} shows 
%
%
%
$g_2>2$ as a function of  $2\kappa/\gamma_{\perp}$ for different   $N_{th}$.     
Fig.~\ref{Fig_5} shows $g_2$ as a function of the dimensionless pump $P$ for various $2\kappa/\gamma_{\perp}$ and $N_{th}$. 
%
%
%
We see the increase of the maximum $g_2$ at $P\rightarrow 0$ with $2\kappa/\gamma_{\perp}$, the decrease with $N_{th}$, and decrease of $g_2(P)$ with $P$. Such a decrease is due to the relative population fluctuation dispersion decrease with $P$: $\delta^2N_e/N_e=1/(P+1)$.  $g_2(P)$  for large $P\geq 1$ can be found using the nonlinear equations~\rf{MBE_0}, taking into account the population fluctuations and the stimulated emission to the laser mode. The population fluctuation spectrum $\delta^2N_e(\omega)$ can be found analytically for arbitrary $P$ using the perturbation procedure described in \ct{PhysRevA.105.053713}. 
\section{Discussion of results\label{Section_5}}
We see from Eq.~\rf{mean_n_norm} that the mean increase in the number of photons $\Delta_n$ due to population fluctuations is large when $N_{th}$ is small, which implies that the coupling constant between the emitter and the field $\Omega\sim N_{th}^{-1/2}$ is large. 
For the bad cavity, when $2\kappa/\gamma_{\perp}>1$, $N_{th}\leq 1$ implies a  single emitter-field strong coupling, when the single emitter-field interaction is reversible, leading to the change of the emitted field spectral structure, as Rabi splitting in the two-level emitter \ct{C7NR06917K,Oraevsky_1994}. 
In such conditions, the inequality \rf{gamma_par_cond} (and hence formulae  \rf{mean_n_norm} and \rf{g2_fin}) is not valid because  $\gamma_{\parallel}$ includes the spontaneous emission rate to the lasing mode. Such a rate is  $\sim\gamma_{\perp}$ for $N_{th}\leq 1$ and $\gamma_{\perp}\leq\kappa$ in SR lasers. 
In Figs.~\ref{Fig_3} and \ref{Fig_4}  $N_{th}\geq 1$ as $N_{th} = 5$, $10$, and $15$, so condition \rf{gamma_par_cond} still holds.

The  increase of $\Delta_n$ (or $n$) by the population fluctuations is possible  in a bad cavity with a large  parameter $2\kappa/\gamma_{\perp}>1$, and significant   collective  effects, as a sub- or superradiance to the cavity mode   \ct{Belyanin_1998,Koch_2017,Khanin}. 
Suppose, we have $N_0$ independent  emitters  and  no collective effects in the radiation. The mean radiation power of the independent emitters is 
$\sim N_e$, so   
fluctuations $\delta \hat{N}_e$ do not contribute to  
$n\sim\left<N_e+\delta \hat{N}_e\right> = N_e$. This is not the case for collective emission, where at least a part of  
$n \sim \left<(N_e+\delta\hat{N}_e)^2\right> = N_e^2+\left<(\delta\hat{N}_e)^2\right>$, which is why the population fluctuations increase 
$n$ with the collective effects.  
 
In our model,  the population fluctuation effect on the radiation causes the superthermal photon statistics.
If we neglect  the population fluctuations, we get $g_2=2$ \ct{Andre:19,Protsenko_2021}. $g_2$ 
increases significantly above $g_2=2$, when $2\kappa/\gamma_{\perp}$ is large and collective effects are significant, in agreement with the results  of \ct{Jahnke,Bohnet,PhysRevA.101.063835}. 
We see that $g_2>2$ is due to $\hat{a}\sim\delta \hat{N}_e$.  
The population fluctuations $\delta \hat{N}_e$ have no phase, so $\left<(\delta\hat{N}_e)^4\right>\sim 3\left<\delta\hat{N}_e^2\right>^2$ for a large number of emitters. 
Otherwise, if we neglect the population fluctuations, then 
$\left<\hat{a}^+\hat{a}^+\hat{a}\hat{a}\right>\sim 2n^2$, leading to $g_2=2$.

We compare our results  with \ct{Jahnke}, where the superthermal photon statistics of 200 Q-dot emitters in a small optical cavity were studied experimentally and  theoretically. 
In our model and in \ct{Jahnke}: (a) the superthermal photon statistics appear at the superradiance in the cavity; (b) when the cavity photon number $n$ is small, $g_2-2 \sim 1$; (c) $g_2>2$ (the photon bunching) increases with a smaller number $N_e$ of emitters in the upper states: $g_2-2\sim \Delta_n^2\sim N_e^{-2}$, see  Eqs.~\rf{mean_n_norm}  and \rf{g2_fin}.  
We do not consider a large $n$ and $g_2$ as in the superradiant pulses studied in \ct{Jahnke}. The superthermal photon statistics found in \ct{Jahnke} for large $n$ may be related to stimulated emission as in \ct{PhysRevLett.52.341, PhysRevA.35.1838}.
In addition to the results of \ct{Jahnke}, we propose that  the population fluctuations are the physical reason for the superthermal photon statistics of the superradiant LEDs at small $n$, and find $g_2>2$ analytically, thanks to our simple two-level active medium model. The model of \ct{Jahnke} is more complicated and realistic. It describes the experiment  through the numerical analysis. 
The population fluctuation effect on the superthermal photon statistics is not mentioned in the paper \ct{Jahnke}.
This may be due to the fact that \ct{Jahnke} uses a four-level model, whereas we consider a two-level model.
The low level in the four-level model is rapidly emptied by the relaxation processes and therefore does not contribute to population fluctuations.
In contrast, both levels in the two-level model contribute to population fluctuations, which result in a "doubling" of the population fluctuation effect in comparison to the four-level model. A significant contribution to the super-thermal photon statistics presented in \ct{Jahnke} may be attributed to the pulsed nature of the signal.
The temporal variability of the photon flux can result in the photon statistics becoming superthermal.   In contrast to \ct{Jahnke}, we consider the stationary radiation regime.  

Superthermal statistics in small superradiant emitters (lasers and LEDs) have been experimentally observed and theoretically described in \ct{Kreinberg2017}. 
$g_2>2$ is found for the emitter with the highest beta factor and the field-matter coupling. 
It confirms our prediction that strong field-matter coupling is favourable for the superthermal photon statistics at superradiant conditions in LEDs and lasers, as it is here in Eq.~\rf{mean_n_norm} with $\Delta_n\sim 1/N_{th}\sim \Omega^2$ ($\Omega$ is the Rabi frequency) and Eq.\rf{g2_fin} with $g_2-2 \sim \Delta_n^2$. 
Similar to our model, the polarisation in \ct{Kreinberg2017} is a dynamic variable, it assumes a bad LED or laser cavity necessary for the superradiance.  

The role of population fluctuation dynamics in superthermal photon statistics has been emphasised in \ct{atoms9010006} (and references therein).
Although our model includes the polarisation of the medium as a dynamical variable (in contrast to the rate equations (7), (8) of \ct{atoms9010006}), the physical interactions in our case are essentially based on the same topological properties as in \ct{atoms9010006}. 
In this light, the coupling of the population fluctuations with the superthermal statistics   does not seem surprising.  

The paper \ct{atoms9010006} points out difficulties related to Langevin terms in the differential equations for quantum LEDs and lasers, such as the negative photon numbers and  numerical instabilities. 
We do not encounter such difficulties with the Langevin force \rf{F_v0} for the weak pump that we consider. However, the difficulties mentioned in \ct{atoms9010006} tell us, that the stability analysis is necessary for high excitations when the nonlinear term $\sim\hat{a}\delta\hat{N}_e$ in  Eqs.~\rf{MBE_0} becomes significant and can lead to instabilities.

Population fluctuations in lasers with a negligibly small superradiance do not lead to the superthermal statistic in the quantum regime as  $n\rightarrow 0$. 
This can be seen, for example, in \ct{10.1063/1.5022958}, where the parameter $2\kappa/\gamma_{\perp} \ll 1$, so that the polarisation is adiabatically eliminated, the superradiance to the lasing mode is negligible, and the rate equation approach \ct{Coldren} is used in the calculations that lead to $g_2\rightarrow 2$ as $n\rightarrow 0$. 
Thus, we propose at least two conditions  necessary for the superthermal photon statistics at $n\rightarrow 0$. 
The first condition is that $2\kappa/\gamma_{\perp} \geq 1$, the polarisation in a dynamical variable, and the superradiance to the lasing mode is non-negligible.
The second condition requires the population fluctuations included in the theory. 
Alternatively, $g_2>2$ can be found with some assumptions in the rate equation (stochastic) approach. 
For example, the difference between the 
rates of spontaneous and stimulated events can be assumed, i.e. the difference between the rates of spontaneous and stimulated emission (and absorption) can be introduced into the rate equations \ct{Andre:20}.
Examples from \ct{Andre:20} and \ct{PhysRevA.101.063835} show that  large fluctuations,  directly observed in a time trace of the photon emission below the laser threshold, can be found in the stochastic approach. 

Our results predict, that a large $g_2$ implies large population fluctuations and that the superthermal photon statistics at $n\rightarrow 0$ are accompanied by the increase in radiation due to the population fluctuations. It will be interesting to experimentally verify  such predictions.

The population fluctuations considered here arise from the rapid and random transitions of emitters from lower to upper states and back due to the pump and population decay.    Another population fluctuation mechanism is the fluctuations in the field-emitter interaction from one emitter to another.  We do not take such fluctuations into account here.   The population fluctuation noise at larger pump $P$ can cause a larger increase of $g_2>2$ below the threshold as in \ct{PhysRevLett.52.341, PhysRevA.35.1838}.
In the future, we will investigate the influence of population fluctuations on the stimulated emission at larger excitation, keeping the nonlinear term $\sim \hat{a}\delta \hat{N}_e$ in Eq.~\rf{MBE_0}, using the Langevin force~\rf{f_v_sum} and the analogue of the cluster expansion method in the
in the frequency domain as described in \ct{PhysRevA.105.053713}.
\section{Conclusion}
We show analytical calculations of the high-order correlations, e.g. in $g_2$, for the quantum two-level laser model, where the field, the polarisation of the active medium and the population are dynamical variables. We find the expression for the polarisation fluctuation Langevin force, with the population fluctuation effect, necessary for the calculations of the high-order correlations and $g_2$. We show that the population fluctuations lead to the superthermal photon statistics $g_2 > 2$ in a strongly quantum case  when the mean photon number $n \rightarrow 0$. In such a case, $g_2 > 2$ requires a large field-emitter coupling and a bad LED cavity with collective effects. Explicit expressions for $g_2$ and $n$, taking into account the population fluctuations, are found under some conditions. The results can be used to understand the statistical properties of quantum radiation, the physics of the operation of small quantum LEDs and lasers with only a few photons in the cavity, and to increase the radiation efficiency of LEDs by the population fluctuations.

The approach to finding high-order quantum correlations can be used in studies of various quantum optical devices. It can be easily generalised to more complicated and realistic laser models with multilevel active medium, inhomogeneous broadening, many modes, etc. The method contributes
to the theory of quantum dissipative systems with resonant oscillations. We hope that our work will be
helpful for experimental studies of the population fluctuation effect on superthermal photon
statistics and radiation in small LEDs and lasers in the quantum regime. 
\appendix
\section{\label{LF_FN}Correlation properties of the Langevin force $\hat{F}^{(N)}_v$ } 
We perform the Fourier expansion of the operators in Eqs.~\rf{MBE_0}  
\beq
\hat{\alpha}(t) = \frac{1}{\sqrt{2\pi}}\int_{-\infty}^{\infty}\hat{\alpha}(\omega)e^{-i\omega t}d\omega, \hspace{0.5cm}\hat{\alpha}^+(t) = \frac{1}{\sqrt{2\pi}}\int_{-\infty}^{\infty}(\hat{\alpha}^+)_{\omega}e^{-i\omega t}d\omega, \lb{F_dec}
\eeq
Here $\hat{\alpha} = \hat{a},\hat{v},...$ and other operators from Eqs.~\rf{MBE_0}, $\hat{\alpha}(\omega)$, ($(\hat{\alpha}^+)_{\omega}$) are  Fourier components of $\hat{\alpha}(t)$ ($\hat{\alpha}^+(t)$) operators and  $(\hat{\alpha}^+)_{\omega} = \hat{\alpha}^+(-\omega)$. The frequency $\omega$ is a deviation from the optical carrier  frequency $\omega_0$ for the operators $\hat{a}$, $\hat{v}$, $\hat{a}_{in}$ and $\hat{F}_v$. Due to $\omega \ll \omega_0$, we replace the low integration limit in Eqs.~\rf{F_dec}  by $-\infty$, which is a common approximation in quantum optics \ct{Scully}. Then we find  algebraic equations for Fourier component operators  
\begin{subequations}\lb{FC_0}\beqr
0&=&\left( i\omega -\kappa  \right)\hat{a}(\omega )+\Omega \hat{v}(\omega )+{{{\hat{F}}}_{a}}(\omega ) \lb{FC_1}\\ 
 0&=&\left( i\omega -{{\gamma }_{\bot }}/2 \right)\hat{v}(\omega )+ \Omega f\left[ \hat{a}(\omega )N+2{{\left( \hat{a}\delta {{{\hat{N}}}_{e}} \right)}_{\omega }} \right]+{{{\hat{F}}}_{v}^{(0)}}(\omega )+{{{\hat{F}}}_{v}^{(N)}}(\omega ).\lb{FC_2}\\
0&=&(i\omega-\gamma_P)\delta\hat{N}_e(\omega)-\Omega \delta\hat{\Sigma }(\omega)+\hat{F}_{N_e}(\omega).  \lb{FC_3} 
\eeqr\end{subequations}
We derive from Eqs.~\rf{FC_1}, \rf{FC_2} taking into account the expression~\rf{f_v_expr} for $\hat{F}_v^{(N)}$
\beq
\hat{a}(\omega )=\hat{a}_0(\omega )+\frac{\kappa {{\gamma }_{\bot }}}{{N}_{th}s(\omega )}\hat{d}(\omega), \hspace{0.5cm} 
 \hat{v}(\omega )=\hat{v}_0(\omega )+\frac{2f\Omega\left( \kappa -i\omega\right)}{s(\omega)}\hat{d}(\omega)\lb{a_FC}
\eeq
where the Fourier component  $\hat{d}(\omega) = \left[ \left( \hat{a}+\hat{b}_{in}^+ \right)\delta {\hat{N}_e} \right]_{\omega }$ and $\hat{a}_0(\omega )$, $\hat{v}_0(\omega )$ are Fourier components of the solutions of Eqs.~\rf{MBE_1}, \rf{MBE_2} with $\delta^2{N}_e = 0$
\begin{subequations}\lb{MBE_fc}\beqr
\hat{a}_0(\omega )&=&[\left( {{\gamma }_{\bot }}/2 -i\omega\right){{{\hat{F}}}_{a}}(\omega )+\Omega {{{\hat{F}}}_{v}}^{(0)}(\omega ) ]/s(\omega )\lb{MBE_fc_1}\\
 \hat{v}_0(\omega )&=&[\left( \kappa -i\omega\right){{{\hat{F}}}_v^{(0)}}(\omega )+\Omega fN\hat{F}_a(\omega ) ]/s(\omega ) \lb{MBE_fc_2}
\eeqr\end{subequations}
where 
\beq
s(\omega)=\left( i\omega -\kappa  \right)\left( i\omega -{{\gamma }_{\bot }}/2 \right)-(\kappa\gamma_{\perp}/2)N/N_{th} \lb{s_small}
\eeq
and $N_{th} = \kappa\gamma_{\perp}/2f\Omega^2$.
Using Eqs.~\rf{MBE_fc} and  the zero order approximation of the Langevin force  \rf{F_v0}  we see that
\beq
\left<[\hat{a}_0(\omega),\hat{a}_0^+(\omega')]\right> = c(\omega)\delta(\omega+\omega'), \lb{comm_rel_pres_0}
\eeq
where $c(\omega)$ is a  commutator spectrum given by Eq.~\rf{comm_sp}. The integral 
$(2\pi)^{-1}\int_{-\infty}^{\infty}c(\omega)d\omega =1$
implies correct Bose commutation relations $[\hat{a}_0(t),\hat{a}_0^+(t)] = 1$. Bose commutation relations must  also hold for the operator $\hat{a}(t)$
\beq
[\hat{a}(t),\hat{a}^+(t)] = 1. \lb{BCR_0}
\eeq
We use  Eq.~\rf{BCR_0} for finding $\hat{b}_{in}$. Taking $\hat{a}(\omega)$ given by Eq.~\rf{a_FC}, we  write the Fourier component commutator
\beq
\left<[\hat{a}(\omega), (\hat{a}^+)_{\omega'}]\right> = 
c(\omega)\delta(\omega+\omega') +\theta(\omega,\omega')\lb{FC_comm}
\eeq
where
\beq
\theta(\omega,\omega')=\frac{\kappa\gamma_{\perp}}{N_{th}}
\left\{\frac{\left<[\hat{d}(\omega),(\hat{a}_0^+)_{\omega'}]\right>}{s(\omega)}
+ \frac{\left<[\hat{a}_0(\omega),(\hat{d}^+)_{\omega'}]\right>}{s^*(-\omega')}\right\}
+\left(\frac{\kappa\gamma_{\perp}}{N_{th}}\right)^2\frac{\left<[\hat{d}(\omega),(\hat{d}^+)_{\omega'}]\right>}{s(\omega)s^*(-\omega')}. \lb{theta_expr}
\eeq
We integrate Eq.~\rf{FC_comm} over $d\omega$ and $d\omega'$, taking into account  $(2\pi)^{-1}\int_{-\infty}^{\infty}c(\omega)d\omega = 1$, and see that the Bose commutation relations \rf{BCR_0} are satisfied if we find $\hat{b}_{in}$ such that $\theta(\omega,\omega') = 0$. Suppose we found such a $\hat{b}_{in}$. Then  
\beq
\left<[\hat{a}_0(\omega), (\hat{a}_0^+)_{\omega'}]\right> = 
\left<[\hat{a}(\omega), (\hat{a}^+)_{\omega'}]\right> = 
c(\omega)\delta(\omega+\omega') \lb{FC_comm_tr}
\eeq
so that the commutator spectrum $c(\omega)$ does not depend on the population fluctuations. 

Let us find $\hat{b}_{in}$ such that $\theta(\omega,\omega') = 0$. The first term in Eq.~\rf{theta_expr} is zero, since  this term is proportional to the first power of the population fluctuations and $\left<\delta \hat{N}_e(\omega)\right>=0$. Therefore $\theta(\omega,\omega')=0$ if \newline $\left<[\hat{d}(\omega),(\hat{d}^+)_{\omega'}]\right>=0$. By examining the second term in Eq.~\rf{theta_expr}, using the relation \rf{FC_comm_tr}, and considering that only  the binary products $\left<\delta\hat{N}_e(\omega)\delta\hat{N}_e(\omega')\right>$ contribute to this term, we find 
\beq
\left<[\hat{d}(\omega),(\hat{d}^+)_{\omega'}]\right> = 
\left\{(c-\left<[\hat{b}_{in},\hat{b}_{in}^+]\right>)*\delta^2N_e\right\}_{\omega}\delta(\omega+\omega'),\lb{diff_cb}
\eeq
where $\{A*B\}_{\omega}$ denotes the Fourier component of the convolution of $A$ and $B$. We  consider that $\hat{b}_{in}(\omega)$ is the Fourier component of the vacuum bath Bose-operator, so $\left<\hat{b}_{in}^+\hat{b}_{in}\right> = 0$. $\hat{b}_{in}$ must satisfy Eq.~\rf{diff_cb} and the Bose commutation relations $[\hat{b}_{in}(t),\hat{b}_{in}^+(t)] = 1$, which is true if $\hat{b}_{in}$ has the spectra \rf{b_in_PS} shown in the main text. 
\section{\label{g2_der}Derivation of the formula for $g_2$ }
Consider 
\[
A=\left<{{\left( {{{\hat{b}}}_{in}}^+ \delta {{{\hat{N}}}_{e}} \right)}_{\omega_1 }}
{{\left( {{{\hat{b}}}_{in}}^+ \delta {{{\hat{N}}}_{e}} \right)}_{\omega_2 }}
{{\left( {{{\hat{b}}}_{in}} \delta {{{\hat{N}}}_{e}} \right)}_{\omega_3 }}{{\left( {{{\hat{b}}}_{in}} \delta {{{\hat{N}}}_{e}} \right)}_{\omega_4 }}\right>,
\]
which appeared in the first integral in Eq.~\rf{cumulant_g}.  The Fourier component of the operator product is a convolution, for example, 
\[\left( \hat{b}_{in}^+\delta \hat{N}_e \right)_{\omega_1}=\frac{1}{\sqrt{2\pi }}\int\limits_{-\infty }^{\infty }{{{\left( {{{\hat{b}_{in}}}^{+}} \right)}_{\omega }}\delta {{{\hat{N}}}_{e}}\left( {{\omega }_{1}}-\omega  \right)d\omega }\equiv\frac{1}{\sqrt{2\pi }}\int\limits_{-\infty }^{\infty }{{{{\hat{b}_{in}}}^{+}}(-\omega )\delta {{{\hat{N}}}_{e}}\left( {{\omega }_{1}}-\omega  \right)d\omega }\]
therefore
\beq A=\frac{1}{{{\left( 2\pi  \right)}^{2}}}\int\limits_{-\infty }^{\infty }{d{{\omega }_{1}}'d{{\omega }_{2}}'d{{\omega }_{3}}'d{{\omega }_{4}}'}F(\omega_1,\omega_2,\omega_3,\omega_4)\lb{A_2}\eeq
with
\beq F=\left< \hat{b}_{in}^{+}(-{{\omega }_{1}}')\hat{b}_{in}^{+}(-{{\omega }_{2}}')\hat{b}_{in}({{\omega }_{3}}')\hat{b}_{in}({{\omega }_{4}}')\right>\cdot \lb{function_F_1}\eeq\[\left<\delta {{{\hat{N}}}_{e}}({{\omega }_{1}}-{{\omega }_{1}}')\delta {{{\hat{N}}}_{e}}({{\omega }_{2}}-{{\omega }_{2}}')\delta {{{\hat{N}}}_{e}}(-{{\omega }_{1}}-{{\omega }_{3}}')\delta {{{\hat{N}}}_{e}}(-{{\omega }_{2}}-{{\omega }_{4}}') \right>. \]
In Eq.~\rf{function_F_1} we use that $\hat{b}_{in}$ and $\delta{{{\hat{N}}}_{e}}$ are uncorrelated and commute. We neglect the rare case, when all four arguments in the first multiplier in Eq.~\rf{function_F_1} are the equal. Therefore
\beq
\left< \hat{b}_{in}^{+}(-{{\omega }_{1}}')\hat{b}_{in}^{+}(-{{\omega }_{2}}')\hat{b}_{in}({{\omega }_{3}}')\hat{b}_{in}({{\omega }_{4}}')\right> = \lb{f_calc_1}\eeq\[c({{\omega }_{3}}')c({{\omega }_{4}}')\delta ({{\omega }_{1}}'+{{\omega }_{3}}')\delta ({{\omega }_{2}}'+{{\omega }_{4}}')+c({{\omega }_{3}}')c({{\omega }_{4}}')\delta ({{\omega }_{1}}'+{{\omega }_{4}}')\delta ({{\omega }_{2}}'+{{\omega }_{3}}').
\]
We insert Eq.~\rf{f_calc_1} into Eq.~\rf{function_F_1}, then Eq.~\rf{function_F_1} into Eq.~\rf{A_2}, calculate integrals over $d\omega_1'$, and $d\omega_2'$ in Eq.~\rf{A_2}  using delta functions, and find instead of Eq.~\rf{A_2}
\beq
A=\frac{2}{{{\left( 2\pi  \right)}^{2}}}\int\limits_{-\infty }^{\infty }{d{{\omega }_{3}}'d{{\omega }_{4}}'}c({{\omega }_{3}}')c({{\omega }_{4}}')\left\langle \delta {{{\hat{N}}}_{e}}({{\omega }_{1}}+{{\omega }_{3}}')\delta {{{\hat{N}}}_{e}}({{\omega }_{2}}+{{\omega }_{4}}')\delta {{{\hat{N}}}_{e}}({{\omega }_{3}}-{{\omega }_{3}}')\delta {{{\hat{N}}}_{e}}({{\omega }_{4}}-{{\omega }_{4}}') \right\rangle \lb{A_exp_22}
\eeq
Note that there are two terms in Eq.~\rf{A_2}, they give the same contributions that are combined in the  expression \rf{A_exp_22}.  We express the mean value in Eq.~\rf{A_exp_22} by the spectrum $\delta^2N_e(\omega)$
\[\left\langle \delta {{{\hat{N}}}_{e}}(\omega )\delta {{{\hat{N}}}_{e}}(\omega ') \right\rangle ={{\delta }^{2}}{{N}_{e}}(\omega )\delta (\omega +\omega ').\]
We consider that $\delta {{{\hat{N}}}_{e}}(-\omega ) = \delta{{{\hat{N}}}_{e}}^+(-\omega )$, and find
\[
  \left\langle \delta {{{\hat{N}}}_{e}}({{\omega }_{1}}+{{\omega }_{3}}')\delta {{{\hat{N}}}_{e}}({{\omega }_{2}}+{{\omega }_{4}}')\delta {{{\hat{N}}}_{e}}({{\omega }_{3}}-{{\omega }_{3}}')\delta {{{\hat{N}}}_{e}}({{\omega }_{4}}-{{\omega }_{4}}') \right\rangle = \]\[ 
 {{\delta }^{2}}{{N}_{e}}({{\omega }_{1}}+{{\omega }_{3}}'){{\delta }^{2}}{{N}_{e}}({{\omega }_{3}}-{{\omega }_{3}}')\delta ({{\omega }_{1}}+{{\omega }_{3}}'+{{\omega }_{2}}+{{\omega }_{4}}')\delta ({{\omega }_{3}}-{{\omega }_{3}}'+{{\omega }_{4}}-{{\omega }_{4}}')+ 
 \]\beq
 {{\delta }^{2}}{{N}_{e}}({{\omega }_{1}}+{{\omega }_{3}}'){{\delta }^{2}}{{N}_{e}}({{\omega }_{2}}+{{\omega }_{4}}')\delta ({{\omega }_{1}}+{{\omega }_{3}})\delta ({{\omega }_{2}}+{{\omega }_{4}})+ \lb{exp_N_e}
 \eeq\[
 {{\delta }^{2}}{{N}_{e}}({{\omega }_{1}}+{{\omega }_{3}}'){{\delta }^{2}}{{N}_{e}}({{\omega }_{2}}+{{\omega }_{4}}')\delta ({{\omega }_{1}}+{{\omega }_{3}}'+{{\omega }_{4}}-{{\omega }_{4}}')\delta ({{\omega }_{2}}+{{\omega }_{4}}'+{{\omega }_{3}}-{{\omega }_{3}}') \]
We insert \rf{exp_N_e} to \rf{A_exp_22}, then \rf{A_exp_22} to \rf{cumulant_g} and integrate in \rf{cumulant_g} over frequencies without primes. These frequencies only appear  in the delta-functions in \rf{exp_N_e}. After the integration the first term in \rf{exp_N_e} gives ${{\omega }_{2}}=-{{\omega }_{1}}-{{\omega }_{3}}'-{{\omega }_{4}}'$ and ${{\omega }_{4}}=-{{\omega }_{3}}+{{\omega }_{3}}'+{{\omega }_{4}}'$; the second term gives $\omega_3=-\omega_1$ and $\omega_4=-\omega_2$; the third term gives ${{\omega }_{4}}=-{{\omega }_{1}}-{{\omega }_{3}}'+{{\omega }_{4}}'$ and ${{\omega }_{3}}=-{{\omega }_{2}}-{{\omega }_{4}}'+{{\omega }_{3}}'$. Therefore in Eq.~\rf{cumulant_g}
\beq\lb{cumulant_calcul}\begin{split}
& \frac{1}{(2\pi)^2}\int_{-\infty}^{\infty}\left<(\hat{B}^+)_{\omega_1}(\hat{B}^+)_{\omega_2}\hat{B}(\omega_3)\hat{B}(\omega_4)\right>d\omega_1...d\omega_4 = \\
  & \frac{2}{{{\left( 2\pi  \right)}^{4}}}\int\limits_{-\infty }^{\infty }{\frac{c({{\omega }_{3}}')c({{\omega }_{4}}'){{\delta }^{2}}{{N}_{e}}({{\omega }_{1}}+{{\omega }_{3}}'){{\delta }^{2}}{{N}_{e}}({{\omega }_{3}}-{{\omega }_{3}}')}{s({{\omega }_{1}})s(-{{\omega }_{1}}-{{\omega }_{3}}'-{{\omega }_{4}}')s({{\omega }_{3}})s(-{{\omega }_{3}}+{{\omega }_{3}}'+{{\omega }_{4}}')}d{{\omega }_{1}}d{{\omega }_{3}}d{{\omega }_{3}}'d{{\omega }_{4}}'}+ \\
 & \frac{2}{{{\left( 2\pi  \right)}^{4}}}\int\limits_{-\infty }^{\infty }{\frac{c({{\omega }_{3}}')c({{\omega }_{4}}'){{\delta }^{2}}{{N}_{e}}({{\omega }_{1}}+{{\omega }_{3}}'){{\delta }^{2}}{{N}_{e}}({{\omega }_{2}}+{{\omega }_{4}}')}{{{\left| s({{\omega }_{1}}) \right|}^{2}}{{\left| s({{\omega }_{2}}) \right|}^{2}}}d{{\omega }_{1}}d{{\omega }_{2}}d{{\omega }_{3}}'d{{\omega }_{4}}'}+ \\ 
 & \frac{2}{{{\left( 2\pi  \right)}^{4}}}\int\limits_{-\infty }^{\infty }{\frac{c({{\omega }_{3}}')c({{\omega }_{4}}'){{\delta }^{2}}{{N}_{e}}({{\omega }_{1}}+{{\omega }_{3}}'){{\delta }^{2}}{{N}_{e}}({{\omega }_{2}}+{{\omega }_{4}}')}{s({{\omega }_{1}})s({{\omega }_{2}})s(-{{\omega }_{2}}-{{\omega }_{4}}'+{{\omega }_{3}}')s(-{{\omega }_{1}}-{{\omega }_{3}}'+{{\omega }_{4}}')}d{{\omega }_{1}}d{{\omega }_{2}}d{{\omega }_{3}}'d{{\omega }_{4}}'} \\ 
\end{split}\eeq
We see that the term proportional to $1/{{\left| s({{\omega }_{1}}) \right|}^{2}}{{\left| s({{\omega }_{2}}) \right|}^{2}}$ under the integral in \rf{cumulant_calcul} is cancelled with the second term in Eq.~\rf{cumulant_g} for the cumulant $C_g$. Then we replace $\omega_2$ by $\omega_3$ and $\omega_4'$ by $-\omega_4'$ in the third integral in Eq.~\rf{cumulant_calcul} taking into account that $c(-\omega_4')=c(-\omega_4')$ and see that the first and the third integrals in \rf{cumulant_calcul} are the same. Therefore
\beq
C_g = \frac{4}{{{\left( 2\pi  \right)}^{4}}}\int\limits_{-\infty }^{\infty }{\frac{c({{\omega }_{3}}')c({{\omega }_{4}}'){{\delta }^{2}}{{N}_{e}}({{\omega }_{1}}+{{\omega }_{3}}'){{\delta }^{2}}{{N}_{e}}({{\omega }_{3}}-{{\omega }_{3}}')}{s({{\omega }_{1}})s(-{{\omega }_{1}}-{{\omega }_{3}}'-{{\omega }_{4}}')s({{\omega }_{3}})s(-{{\omega }_{3}}+{{\omega }_{3}}'+{{\omega }_{4}}')}d{{\omega }_{1}}d{{\omega }_{3}}d{{\omega }_{3}}'d{{\omega }_{4}}'}. \lb{C_g_interm_res}
\eeq
Make the substitutions $\tilde{\omega}_1 = {\omega}_1+{\omega}_3'$, $\tilde{\omega}_3 = {\omega}_3-{\omega}_3'$ instead of ${\omega}_1$ and ${\omega}_3$, respectively; then replacing  $\omega_4'$ by $-\omega_4'$ and $\tilde{\omega}_1$ by $-\tilde{\omega}_1$; using $c(\omega) = c(-\omega)$, $\delta^2N_e(\omega) = \delta^2N_e(-\omega)$ and $s(-\omega) = s^*(\omega)$ we represent \rf{C_g_interm_res} as 
\beq
C_g = \frac{4}{(2\pi)^2}\int_{-\infty}^{\infty}c(\omega_3')c(\omega_4')|J_g(\omega_3',\omega_4')|^2d\omega_3'd\omega_4' \lb{C_g_interm_res_1}
\eeq
where
\beq
J_g(\omega_3',\omega_4') = \frac{1}{2\pi}\int_{-\infty}^{\infty}\frac{\delta^2N_e(\omega)d\omega}{s(\omega+\omega_4')s^*(\omega+\omega_3')}. \lb{j_g_int}
\eeq
Eq.~\rf{C_g_interm_res_1} is simplified at the condition \rf{gamma_par_cond}, which allows to approximate $\delta^2N_e(\omega) \approx \delta^2N_e\delta(\omega)$ in Eq.~\rf{j_g_int}, then
\beq
C_g = \delta^2N_e\left[\frac{2}{2\pi}\int_{-\infty}^{\infty}\frac{c(\omega)}{|s(\omega)|^2}d\omega\right]^2 \lb{C_g_interm_res_2}
\eeq
and therefore, according to Eq.~\rf{g_2_cummul},
\beq
g_2 = 2 + \left(\frac{\kappa\gamma_{\perp}}{N_{th}}\right)^4\frac{\delta^2N_e}{n^2}\left[\frac{2}{2\pi}\int_{-\infty}^{\infty}\frac{c(\omega)}{|s(\omega)|^2}d\omega\right]^2 = 2\left[1 + 2\left(\frac{\Delta n}{n_0+\Delta n}\right)^2\right], \lb{g_2_cummul_app}
\eeq
where we represent $n=n_0+\Delta n$ and $\Delta n$ is a part of $n$, given by Eq.~\rf{n_pop_fl_0},  caused by the population fluctuations. Taking $\Delta_n = \Delta n/n_0$ in Eq.~\rf{g_2_cummul_app}, we obtain Eq.~\rf{g2_fin} of the main text.

\medskip

\bibliographystyle{MSP}
\bibliography{myrefs}

\begin{thebibliography}{10}
\providecommand{\url}[1]{\texttt{#1}}
\providecommand{\urlprefix}{URL }

\bibitem{Huang:16}
C.-H. Huang, Y.-H. Wen, Y.-W. Liu,
\newblock \emph{Opt. Express} \textbf{2016}, \emph{24}, 4 4278.

\bibitem{RevModPhys.70.101}
M.~B. Plenio, P.~L. Knight,
\newblock \emph{Rev. Mod. Phys.} \textbf{1998}, \emph{70} 101.

\bibitem{PhysRevLett.97.083604}
J.~S. Neergaard-Nielsen, B.~M. Nielsen, C.~Hettich, K.~M\o{}lmer, E.~S. Polzik,
\newblock \emph{Phys. Rev. Lett.} \textbf{2006}, \emph{97} 083604.

\bibitem{Michler2000}
P.~Michler, A.~Imamoglu, M.~D. Mason, P.~J. Carson, G.~F. Strouse, S.~K. Buratto,
\newblock \emph{Nature} \textbf{2000}, \emph{406}, 6799 968.

\bibitem{Das:10}
M.~Das, A.~Shirasaki, K.~P. Nayak, M.~Morinaga, F.~L. Kien, K.~Hakuta,
\newblock \emph{Opt. Express} \textbf{2010}, \emph{18}, 16 17154.

\bibitem{PhysRevLett.76.900}
F.~De~Martini, G.~Di~Giuseppe, M.~Marrocco,
\newblock \emph{Phys. Rev. Lett.} \textbf{1996}, \emph{76} 900.

\bibitem{9466840}
T.~Wang, C.~Jiang, J.~Zou, H.~Zhou, X.~Lin, H.~Chen, G.~P. Puccioni, G.~Wang, G.~L. Lippi,
\newblock \emph{IEEE Sensors Journal} \textbf{2021}, \emph{21}, 18 19948.

\bibitem{George:21}
A.~George, A.~Bruhacs, A.~Aadhi, R.~Ostic, E.~Whitby, W.~E. Hayenga, Z.~M. Wang, M.~Kues, C.~Reimer, M.~Khajavikhan, R.~Morandotti,
\newblock In \emph{OSA Advanced Photonics Congress 2021}. Optica Publishing Group, \textbf{2021} IF1A.5.

\bibitem{Hayenga:16}
W.~E. Hayenga, H.~Garcia-Gracia, H.~Hodaei, C.~Reimer, R.~Morandotti, P.~LiKamWa, M.~Khajavikhan,
\newblock \emph{Optica} \textbf{2016}, \emph{3}, 11 1187.

\bibitem{Hertel}
I.~V. Hertel, C.-P. Schulz,
\newblock \emph{Atoms, Molecules and Optical Physics 2 Molecules and Photons - Spectroscopy and Collisions},
\newblock Springer, \textbf{2015}.

\bibitem{Fox2006}
M.~Fox,
\newblock \emph{Quantum optics},
\newblock Oxford University Press, \textbf{2006}.

\bibitem{Jahnke}
F.~Jahnke, C.~Gies, M.~A{\ss}mann, M.~Bayer, H.~A.~M. Leymann, A.~Foerster, J.~Wiersig, C.~Schneider, M.~Kamp, S.~H{\"o}fling,
\newblock \emph{Nature Commun.} \textbf{2016}, \emph{7} 11540.

\bibitem{PhysRevLett.52.341}
P.~Lett, R.~Short, L.~Mandel,
\newblock \emph{Phys. Rev. Lett.} \textbf{1984}, \emph{52} 341.

\bibitem{PhysRevA.35.1838}
R.~F. Fox, R.~Roy,
\newblock \emph{Phys. Rev. A} \textbf{1987}, \emph{35} 1838.

\bibitem{Auffeves_2011}
A.~Auffeves, D.~Gerace, S.~Portolan, A.~Drezet, M.~F. Santos,
\newblock \emph{New Journal of Physics} \textbf{2011}, \emph{13}, 9 093020.

\bibitem{PhysRevA.81.063827}
D.~Meiser, M.~J. Holland,
\newblock \emph{Phys. Rev. A} \textbf{2010}, \emph{81} 063827.

\bibitem{PhysRevApplied.10.054055}
F.~Lohof, R.~Barzel, P.~Gartner, C.~Gies,
\newblock \emph{Phys. Rev. Appl.} \textbf{2018}, \emph{10} 054055.

\bibitem{PhysRevA.99.053820}
N.~Takemura, M.~Takiguchi, E.~Kuramochi, A.~Shinya, T.~Sato, K.~Takeda, S.~Matsuo, M.~Notomi,
\newblock \emph{Phys. Rev. A} \textbf{2019}, \emph{99} 053820.

\bibitem{Kreinberg2017}
S.~Kreinberg, W.~W. Chow, J.~Wolters, C.~Schneider, C.~Gies, F.~Jahnke, S.~H{\"o}fling, M.~Kamp, S.~Reitzenstein,
\newblock \emph{Light Sci. Appl.} \textbf{2017}, \emph{6}, 8 e17030.

\bibitem{PhysRevX.6.011025}
{M. A. Norcia }, J.~K. Thompson,
\newblock \emph{Phys. Rev. X} \textbf{2016}, \emph{6} 011025.

\bibitem{PhysRevA.96.013847}
S.~A. Sch\"affer, B.~T.~R. Christensen, M.~R. Henriksen, J.~W. Thomsen,
\newblock \emph{Phys. Rev. A} \textbf{2017}, \emph{96} 013847.

\bibitem{PhysRevA.81.033847}
{D. Meiser }, M.~J. Holland,
\newblock \emph{Phys. Rev. A} \textbf{2010}, \emph{81} 033847.

\bibitem{PhysRevA.98.063837}
K.~Debnath, Y.~Zhang, K.~M\o{}lmer,
\newblock \emph{Phys. Rev. A} \textbf{2018}, \emph{98} 063837.

\bibitem{Bohnet}
J.~G. Bohnet, Z.~Chen, J.~M. Weiner, D.~Meiser, M.~J. Holland, J.~K. Thompson,
\newblock \emph{Nature} \textbf{2012}, \emph{484} 78.

\bibitem{Bhatti2015}
D.~Bhatti, J.~von Zanthier, G.~S. Agarwal,
\newblock \emph{Scientific Reports} \textbf{2015}, \emph{5}, 1 17335.

\bibitem{Blazek:11}
M.~Blazek, S.~Hartmann, A.~Molitor, W.~Elsaesser,
\newblock \emph{Opt. Lett.} \textbf{2011}, \emph{36}, 17 3455.

\bibitem{Kreinberg_01}
S.~Kreinberg, K.~Laiho, F.~Lohof, W.~E. Hayenga, P.~Holewa, C.~Gies, M.~Khajavikhan, S.~Reitzenstein,
\newblock \emph{Laser \& Photonics Reviews} \textbf{2020}, \emph{14}, 12 2000065.

\bibitem{PhysRevLett.130.253801}
M.~Bundgaard-Nielsen, E.~V. Denning, M.~Saldutti, J.~M\o{}rk,
\newblock \emph{Phys. Rev. Lett.} \textbf{2023}, \emph{130} 253801.

\bibitem{PhysRev.185.568}
M.~Lax, W.~H. Louisell,
\newblock \emph{Phys. Rev.} \textbf{1969}, \emph{185} 568.

\bibitem{PhysRevApplied.4.044018}
H.~A.~M. Leymann, A.~Foerster, F.~Jahnke, J.~Wiersig, C.~Gies,
\newblock \emph{Phys. Rev. Appl.} \textbf{2015}, \emph{4} 044018.

\bibitem{10.1063/1.5138937}
M.~Sanchez-Barquilla, R.~E.~F. Silva, J.~Feist,
\newblock \emph{The Journal of Chemical Physics} \textbf{2020}, \emph{152}, 3 034108.

\bibitem{Leymann_01}
H.~A.~M. Leymann, A.~Foerster, J.~Wiersig,
\newblock \emph{physica status solidi c} \textbf{2013}, \emph{10}, 9 1242.

\bibitem{Plankensteiner2022quantumcumulantsjl}
D.~Plankensteiner, C.~Hotter, H.~Ritsch,
\newblock \emph{{Quantum}} \textbf{2022}, \emph{6} 617.

\bibitem{PhysRevA.59.1667}
I.~Protsenko, P.~Domokos, V.~Lef\`evre-Seguin, J.~Hare, J.~M. Raimond, L.~Davidovich,
\newblock \emph{Phys. Rev. A} \textbf{1999}, \emph{59} 1667.

\bibitem{Andre:19}
E.~C. Andr\'{e}, I.~E. Protsenko, A.~V. Uskov, J.~M{\o}rk, M.~Wubs,
\newblock \emph{Opt. Lett.} \textbf{2019}, \emph{44}, 6 1415.

\bibitem{Protsenko_2021}
I.~E. Protsenko, A.~V. Uskov, E.~C. Andr{\'{e}}, J.~M{\o}rk, M.~Wubs,
\newblock \emph{New Journal of Physics} \textbf{2021}, \emph{23}, 6 063010.

\bibitem{PhysRevA.105.053713}
I.~E. Protsenko, A.~V. Uskov,
\newblock \emph{Phys. Rev. A} \textbf{2022}, \emph{105} 053713.

\bibitem{https://doi.org/10.1002/andp.202200298}
I.~E. Protsenko, A.~V. Uskov,
\newblock \emph{Annalen der Physik} \textbf{2023}, \emph{535}, 1 2200298.

\bibitem{PhysRevA.18.1628}
F.~Casagrande, R.~Cordoni,
\newblock \emph{Phys. Rev. A} \textbf{1978}, \emph{18} 1628.

\bibitem{PhysRevA.101.063835}
T.~Wang, D.~Aktas, O.~Alibart, E.~Picholle, G.~P. Puccioni, S.~Tanzilli, G.~L. Lippi,
\newblock \emph{Phys. Rev. A} \textbf{2020}, \emph{101} 063835.

\bibitem{Belyanin_1998}
A.~A. Belyanin, V.~V. Kocharovsky, V.~V. Kocharovsky,
\newblock \emph{Quant. Semiclass. Opt.: JEOS Part B} \textbf{1998}, \emph{10}, 2 L13.

\bibitem{Koch_2017}
V.~V. Kocharovsky, V.~V. Zheleznyakov, E.~R. Kocharovskaya, V.~V. Kocharovsky,
\newblock \emph{Physics-Uspekhi} \textbf{2017}, \emph{60}, 4 345.

\bibitem{Khanin}
Y.~I. Khanin,
\newblock \emph{Fundamentals of laser dynamics},
\newblock Cambridge International Science Pub, \textbf{2005}.

\bibitem{Scully}
M.~S. Scully, M. O.~Zubairy,
\newblock \emph{Quantum {O}ptics},
\newblock Cambridge University Press, \textbf{1997}.

\bibitem{Yariv}
A.~Yariv,
\newblock \emph{Quantum {E}lectronics},
\newblock John Wiley and Sons, Inc., \textbf{1967}.

\bibitem{1071726}
Y.~{Yamamoto},
\newblock \emph{IEEE J. Quant. Electron.} \textbf{1983}, \emph{19}, 1 34.

\bibitem{RevModPhys.68.127}
L.~Davidovich,
\newblock \emph{Rev. Mod. Phys.} \textbf{1996}, \emph{68} 127.

\bibitem{PhysRevA.47.1431}
M.~I. Kolobov, L.~Davidovich, E.~Giacobino, C.~Fabre,
\newblock \emph{Phys. Rev. A} \textbf{1993}, \emph{47} 1431.

\bibitem{Coldren}
L.~A. Coldren, S.~W. Corzine, M.~L. Masanovic,
\newblock \emph{Diode lasers and photonic integrated circuits},
\newblock Wiley, 2nd ed., \textbf{2012}.

\bibitem{10.1063/1.5022958}
J.~Mork, G.~L. Lippi,
\newblock \emph{Applied Physics Letters} \textbf{2018}, \emph{112}, 14 141103.

\bibitem{atoms9010006}
G.~L. Lippi,
\newblock \emph{Atoms} \textbf{2021}, \emph{9}, 1.

\bibitem{Glauber}
R.~J. Glauber,
\newblock In S.~Sarkar, E.~R. Pike, editors, \emph{Frontiers in quantum optics}. Hilger, Boston, MA, \textbf{1986}.

\bibitem{Stenholm_1986}
S.~Stenholm,
\newblock \emph{Physica Scripta} \textbf{1986}, \emph{T12} 56.

\bibitem{C7NR06917K}
D.~S. Dovzhenko, S.~V. Ryabchuk, Y.~P. Rakovich, I.~R. Nabiev,
\newblock \emph{Nanoscale} \textbf{2018}, \emph{10} 3589.

\bibitem{Oraevsky_1994}
A.~N. Oraevsky,
\newblock \emph{Physics-Uspekhi} \textbf{1994}, \emph{37}, 4 393.

\bibitem{Andre:20}
E.~C. Andr\'{e}, J.~M{\o}rk, M.~Wubs,
\newblock \emph{Opt. Express} \textbf{2020}, \emph{28}, 22 32632.

\end{thebibliography}

\end{document}